\documentclass[prc,aps,amsfonts,amsmath,superscriptaddress,floatfix,twocolumn]{revtex4}
\usepackage{graphicx,float} 
\usepackage{epsfig} 
\usepackage{rotating}
\usepackage{mwe}    
\usepackage{subfig}

\begin{document}

\title{Single-particle spatial dispersion and clusters in nuclei}

\author{J.-P. Ebran}
\affiliation{CEA,DAM,DIF, F-91297 Arpajon, France}
\author{E. Khan}
\affiliation{Institut de Physique Nucl\'eaire, Universit\'e Paris-Sud, IN2P3-CNRS, 
Universit\'e Paris-Saclay, F-91406 Orsay Cedex, France}
\author{ R.-D Lasseri}
\affiliation{Institut de Physique Nucl\'eaire, Universit\'e Paris-Sud, IN2P3-CNRS, 
Universit\'e Paris-Saclay, F-91406 Orsay Cedex, France}
\author{D. Vretenar}
\affiliation{Physics Department, Faculty of Science, University of
Zagreb, 10000 Zagreb, Croatia}

\begin{abstract} 
The spatial dispersion of the single-nucleon wave functions is analyzed using the 
self-consistent mean-field framework based on nuclear energy density functionals, 
and with the harmonic oscillator approximation for the nuclear potential. It is shown that 
the dispersion depends on the radial quantum number n, but displays only a very weak dependence on 
the orbital angular momentum. An analytic expression is derived 
for the localization parameter that explicitly takes into account the radial quantum 
number of occupied single-nucleon states. The conditions for single-nucleon localization and 
formation of cluster structures are fulfilled in relatively light nuclei with $A \leq 30$ and $n=1$ states occupied. 
Heavier nuclei exhibit the quantum liquid phase of nucleonic matter because occupied levels that originate from $n > 1$ spherical states are 
largely delocalized. Nevertheless, individual $\alpha$-like clusters can be formed from valence nucleons filling single-particle levels 
originating from $n=1$ spherical mean-field states. 
\end{abstract}
 


\date{\today}

\maketitle


Nucleon localization and formation of cluster structures characterize not only light $\alpha$-conjugate nuclei \cite{Hor12,Kim16,Ita01,Rei11,Toh17,ebr17}, but also heavier nuclear systems \cite{Beck}. Several microscopic models, for instance the anti-symmetrized molecular dynamics \cite{Hor12,Kim16}, have very successfully been applied to a description of cluster states in relatively light nuclei. A more general approach based on energy density functionals (EDFs)
\cite{aru05,ebr12,ebr14,mar18,naza2} has to be employed in order to study the occurrence and structure of nucleon clusters in medium-heavy and heavy nuclei. Studies based 
on the latter method have related the conditions for nucleon localization and formation of clusters to the underlying single-nucleon dynamics, geometric shape 
transitions and surface effects. 

The EDF framework enables a systematic analysis of nucleon localization as a precondition for cluster formation. The Wigner parameter \cite{wig34} can be used to describe the transition between the nuclear quantum liquid phase and a hybrid phase of cluster states in terms of spatial localization. 
Such a localization parameter can be microscopically calculated as the ratio of the dispersion of a single-nucleon wave function to the average inter-nucleon distance \cite{ebr12,ebr13}. If the confining nuclear potential is approximated by a three-dimensional harmonic oscillator (HO), the analytic form of the localization parameter exhibits an explicit dependence on the number of nucleons A and the depth of the confining potential V$_0$. When the spatial dispersion of the single-nucleon 
wave function is of the same size as as the inter-nucleon distance, localization facilitates the formation of clusters \cite{ebr17}. In the present study we aim to explore the dependence of the localization parameter on the specific quantum states occupied by the valence nucleons. A recent study \cite{afa18} has shown explicitly, 
using as examples $^{12}$C, $^{28}$Si and $^{40}$Ca, that single-particle wave functions are localized in light nuclei. Clusters, of course, occur more frequently in light nuclei, but they may also form in heavy systems such as, for instance, an $\alpha$-like cluster in $^{212}$Po \cite{ast10}. Can we understand these phenomena in a unified framework?

If the average nuclear potential is approximated by a spherical harmonic oscillator, one obtains the following analytic expression for the localization parameter  \cite{ebr12,ebr13,ebr14a}
\begin{equation}
\alpha_{loc}\simeq\frac{b}{r_0}=\frac{\sqrt{\hbar}A^{1/6}}{(2mV_0r_0^2)^{1/4}}
\label{adef}
\end{equation}
where $b = \sqrt{\hbar /m \omega_0}$ fm is the oscillator length, and  ${r_0 \simeq 1.25}$ fm is the typical inter-nucleon distance 
determined by nuclear saturation density ($\rho \simeq 0.16$ fm$^{-3})$. The resulting expression includes the 
nucleon number $A$, the mass of the nucleon $m$, and the depth of the confining potential V$_0$. As shown in Ref.~\cite{ebr13}, the oscillator length can be related to the 
spatial dispersion $\Delta r = \sqrt{<r^2> - <r>^2}$: $b \simeq 2 \Delta r$ for the first s, p, and d HO wave functions.
When the dispersion of the single-nucleon wave function is of the same size as the inter-nucleon distance, $\alpha_{loc}$ is of the order of $1$ and
and this facilitates the formation of $\alpha$-clusters. 
The dependence of the localization parameter on $A^{1/6}$ means that 
cluster states are preferably formed in lighter nuclei, and the transition 
from coexisting cluster and mean-field states to a Fermi liquid state should
occur for nuclei with $A \approx 20 - 30$, in qualitative agreement with experiment. In finite nuclei the spatial dispersion and, therefore the localization 
parameter, will explicitly depend on the quantum numbers of specific single-nucleon orbitals. 
In this work we generalize expression (\ref{adef}) and derive an explicit dependence of the localization parameter on single-nucleon quantum numbers. We also compare the spatial dispersions of the HO wave functions with those obtained in a fully self-consistent microscopic calculation of nuclear ground states, and 
perform a systematic microscopic calculation of single-nucleon dispersions in axially symmetric nuclei over the entire nuclear chart.

\begin{figure}[tb]
\scalebox{0.33}{\includegraphics{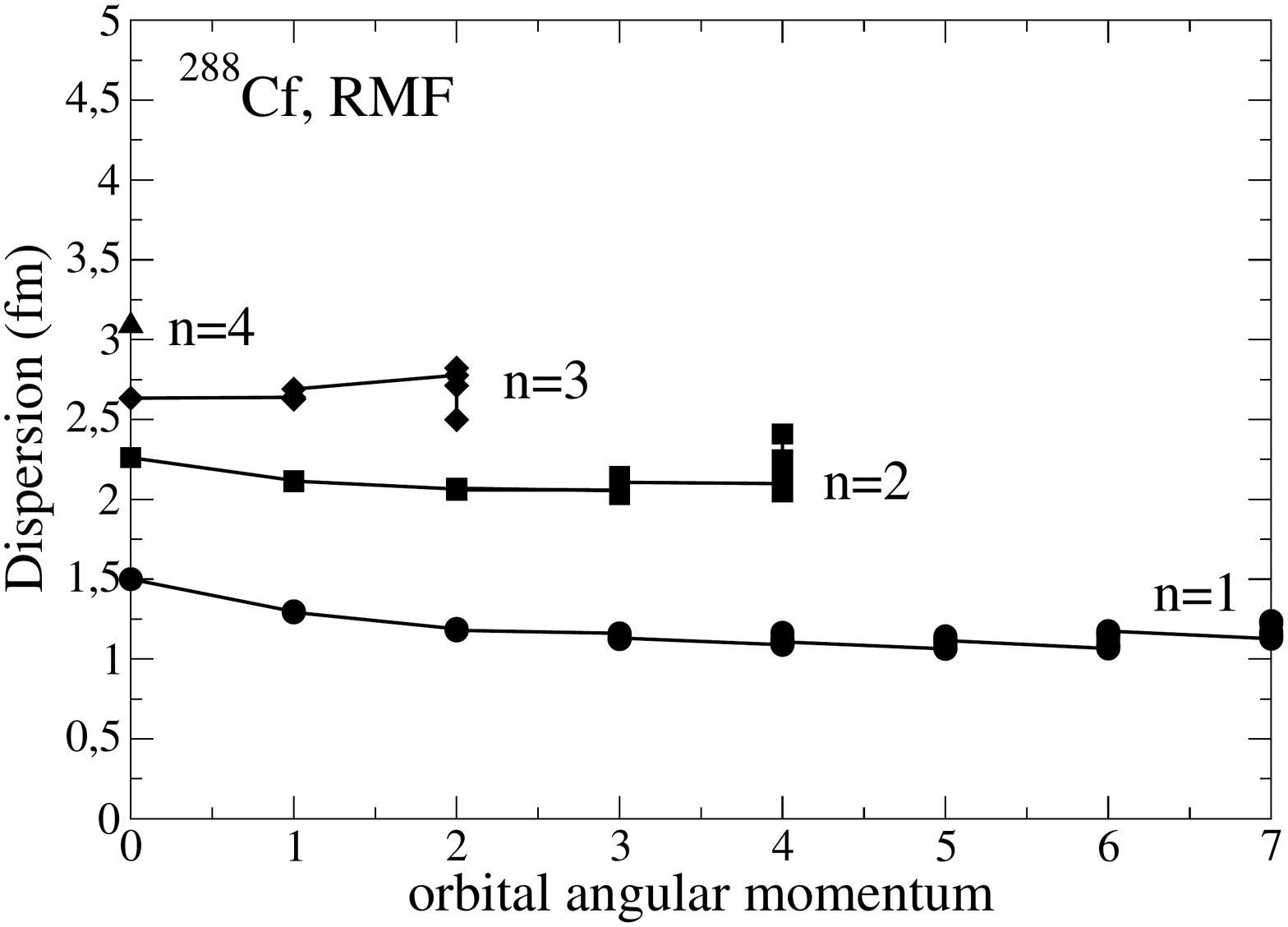}}
 \caption{Radial dispersions $\Delta$r of the single-neutron wave functions of $^{288}$Cf, obtained in a self-consistent mean-field 
 (RMF) calculation based on the  energy density functional DD-ME2 \cite{lala}.}    
 \label{fig:dsm}
\end{figure}

In the first step we perform a systematic microscopic calculation, based on the EDF framework, of dispersions of single-nucleon wave functions in a large nucleus, close to the spherical shape. By considering a heavy spherical nucleus with many occupied levels we can analyze the dependence of the corresponding dispersions on the radial and orbital quantum numbers. 
Figure \ref{fig:dsm} displays the spatial dispersions of neutron single particle states in $^{288}$Cf, obtained in a self-consistent relativistic mean-field (RMF) calculation using the energy density functional DD-ME2 \cite{lala}. The depth of the self-consistent neutron potential is V$_0$=78.6 MeV, and the dispersions $\Delta$r 
are plotted as functions of the single-particle radial quantum number $n$ and orbital angular momentum $l$. One notices a pronounced dependence on the 
radial quantum number $n$, whereas the spatial dispersions $\Delta$r depend only very weakly on the orbital angular momentum. A particularly interesting result is 
that for single-neutron states with $n=1$ the dispersion is of the size of the average inter-nucleon distance. We note that the small splittings between points that correspond to the 
same orbital angular momenta and radial quantum numbers arise because of deformation: the self-consistent mean-field solution is not fully spherical symmetric (the quadrupole 
deformation parameter is $\beta_2 = 0.07$).

\begin{figure}[tb]
\scalebox{0.33}{\includegraphics{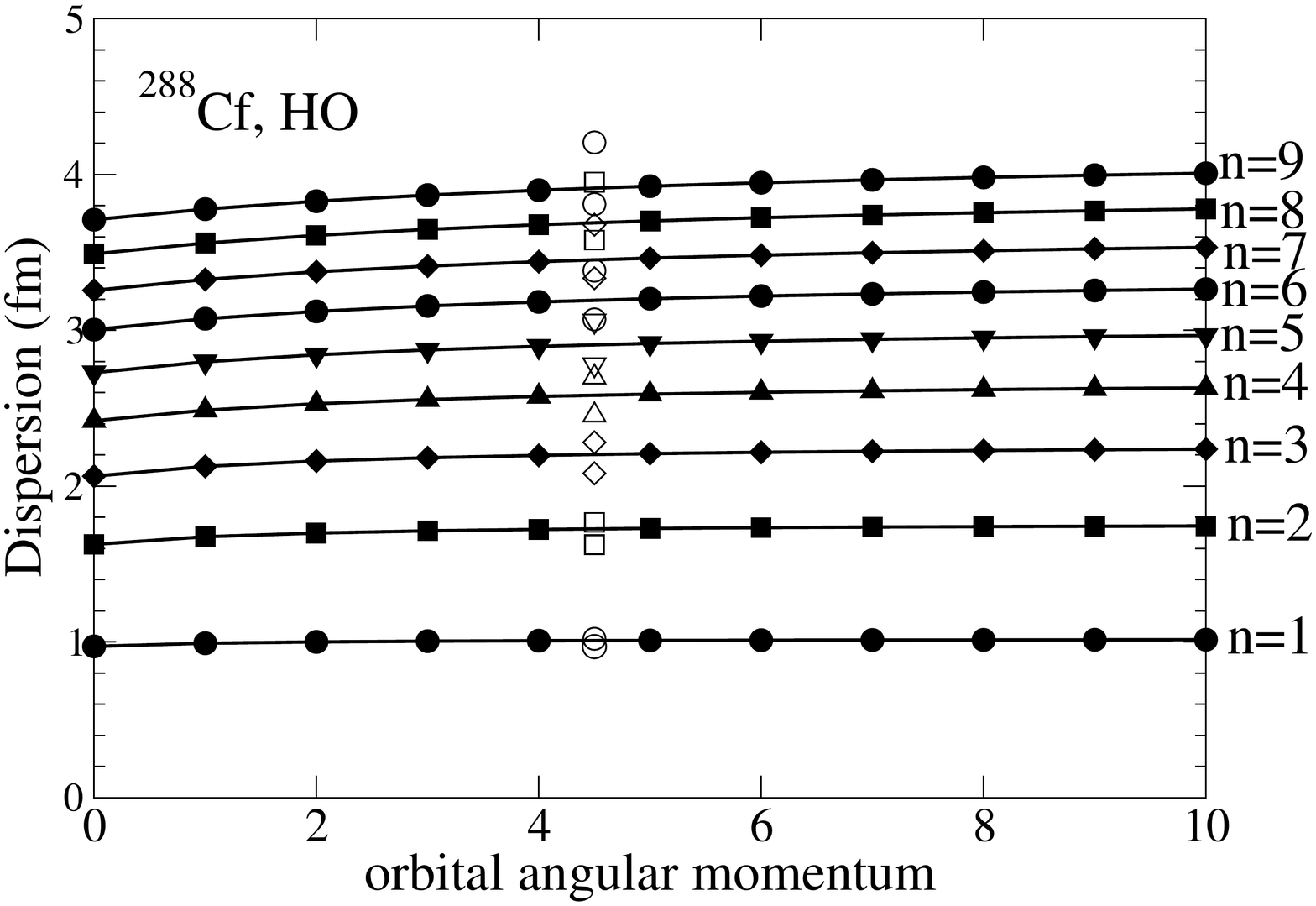}}
 \caption{Radial dispersions $\Delta$r of the harmonic oscillator wave functions of $^{288}$Cf, evaluated numerically from Eqs. (\ref{eq:d2r}) and (\ref{eq:dr}) (filled symbols), and in the analytical approximation 
 (Eqs.~(\ref{eq:d1}) and (\ref{eq:d2}) corresponding to the minimal and maximal values, respectively) (open symbols).}    
 \label{fig:HOn}
\end{figure}

Next we derive an analytic expression for the dispersion of the single-nucleon wave function, for the case when the nuclear potential is approximated by a spherical 
three-dimensional harmonic oscillator. The HO approach provides a realistic approximation for studies of localization and cluster effects in nuclear systems \cite{sch17}. The $<$r$^2>$ term is easy to evaluate and reads:
 \begin{equation}
 <r^2>=b^2\left(N+\frac{3}{2}\right)=b^2\left(2n'+l+\frac{3}{2}\right)
 \label{eq:d2r}
 \end{equation}
where $N=2(n-1) + l$ is the principal quantum number, and $n' \equiv n-1$. 
The $<$r$>$ term is considerably more complicated. Using the HO wave functions, it can be expressed in the following form
 \begin{equation}
\frac{<r>}{b}=\sum_{q=0}^{n'}\frac{(-1)^q(l+q+1)!\Gamma\left(n'-q-\frac{1}{2}\right)}{q!(n'-q)!\Gamma\left(l+q+\frac{3}{2}\right)\Gamma\left(-q-\frac{1}{2}\right)}
\label{eq:dr}
 \end{equation}
where $\Gamma$ is the Euler function.
To compare with the microscopic results shown in Fig. \ref{fig:dsm}, the corresponding dispersions for the single-particle wave functions of the 
 harmonic oscillator potential of $^{288}$Cf are evaluated numerically using Eqs. (\ref{eq:d2r}) and (\ref{eq:dr}), and plotted in Fig.~\ref{fig:HOn}. The dispersion, of course, increases with the number of radial nodes, but shows very little dependence on the orbital angular momentum, just as in the case of a fully microscopic calculation.
 It should be noted that the microscopic dispersion (cf. Fig.~\ref{fig:dsm}) is typically 1.2 times larger than the corresponding one in the HO approximation, because the actual self-consistent nuclear potential is more diffuse. Indeed, a Woods-Saxon potential can be approximated by a HO with a length of about $1.2b$, thus explaining this ratio.

Therefore, if only $n=1$ states are occupied in a nucleus, all nucleons have similar and minimal spatial dispersion, of the order of 1 fm. The pronounced localization 
will favor formation of $\alpha$-like clusters, whereas the occupation of $n > 1$ states breaks the coherence of spatial localization. Of course, nuclei in which only 
levels originating from the $n=1$ spherical states are occupied are the light ones up to about silicon ($Z=14$, $1s,1p,1d$ levels occupied). These are indeed  nuclear systems in which cluster structures are empirically most pronounced \cite{freer}.

To derive a generalization of the expression for the localization parameter in the HO approximation Eq.~(\ref{adef}), 
but now taking explicitly into account the quantum numbers of occupied  
states, we simplify the $n$ and $l$ dependence in Eq.~(\ref{eq:dr}). For $l=0$ one obtains 
\begin{equation}
\frac{<r>}{b}=\frac{2}{\sqrt{\pi}}\frac{(2n'+1)!!}{(2n')!!}\simeq\frac{2}{\sqrt{\pi}}\left(\frac{5n'}{4}+1\right)^{1/2}
\label{eq:drl0}
\end{equation}
 where the $rhs$ is an accurate approximation with a $<$1\% error for $n'=20$. 
Thus, using Eqs. (\ref{eq:d2r}) and (\ref{eq:drl0}), the $l=0$ dispersion reads:
\begin{equation}
\left(\frac{\Delta r}{b}\right)^2\simeq\left(2-\frac{5}{\pi}\right)n'+\left(\frac{3}{2}-\frac{4}{\pi}\right)\simeq 0.4n'+0.23
\label{eq:d1}
\end{equation}
Let us now consider the case of large angular momenta $l$ in Eq.~(\ref{eq:dr}). In this limit \cite{abra}:
 \begin{equation}
 \frac{(l+q+1)!}{\Gamma\left(l+q+\frac{3}{2}\right)}\simeq \sqrt{l}+\frac{1}{\sqrt{l}}\left(\frac{q}{2}+\frac{5}{8}\right)\;,
 \end{equation}
and the expression Eq.~(\ref{eq:dr}) reduces to:
\begin{equation}
\frac{<r>}{b}\simeq \sqrt{l} + \frac{1}{\sqrt{l}}\left(\frac{5}{8}+\frac{3n'}{4}\right)\,.
\label{eq:drlg}
 \end{equation}
The corresponding dispersion for large $l$ values reads:
\begin{equation}
\left(\frac{\Delta r}{b}\right)^2\simeq\frac{n'}{2}+\frac{1}{4}\;.
\label{eq:d2}
\end{equation}

The close agreement of the expressions for $l=0$ (Eq.~(\ref{eq:d1})) and in the large $l$ limit (Eq.~(\ref{eq:d2})), reflects the weak dependence of the HO dispersion 
on orbital angular momentum. The corresponding dispersions of occupied states of $^{288}$Cf: minimal values corresponding to Eq. (\ref{eq:d1}) and maximal 
values computed using Eq.~(\ref{eq:d2}), are indicated by open symbols in Fig.~\ref{fig:HOn}. Both expressions, of course, yield very similar dispersions. 
Eq.~(\ref{eq:d2}) implies, as also shown in Fig.~\ref{fig:HOn}, that the occupation of an $n=2$ state leads to a dispersion that is by a 
factor $\sqrt{3}\sim 1.7$ larger than the one of $n=1$ states. This corresponds to the case of medium-heavy nuclei, typically above silicon, 
in which there is no clear evidence of cluster states at low energy and angular momenta. 

From Eqs.~(\ref{adef}) and (\ref{eq:d2}) we finally derive the approximate expression for the HO localization parameter:
\begin{equation}
\alpha_{loc}\simeq\frac{b}{r_0}\sqrt{2n-1}=\frac{\sqrt{\hbar (2n-1)}}{(2mV_0r_0^2)^{1/4}}~A^{1/6}\;.
\label{eq:res}
\end{equation}
In nuclei the depth of the confining potential is rather constant, as well as the average inter-nucleon distance, hence the two key parameters that determine 
localization are A and the radial quantum number $n$. For relatively light nuclei, with $A \leq 30$ and $n=1$ states occupied, $\alpha_{loc} \lesssim 1$ and 
this favors the formation of $\alpha$-like clusters. In heavier nuclei levels that originate from $n > 1$ spherical states are largely delocalized and this explains the 
predominant liquid drop nature of these systems. 

\begin{figure}[tb]
\begin{center}
\scalebox{0.35}{\includegraphics{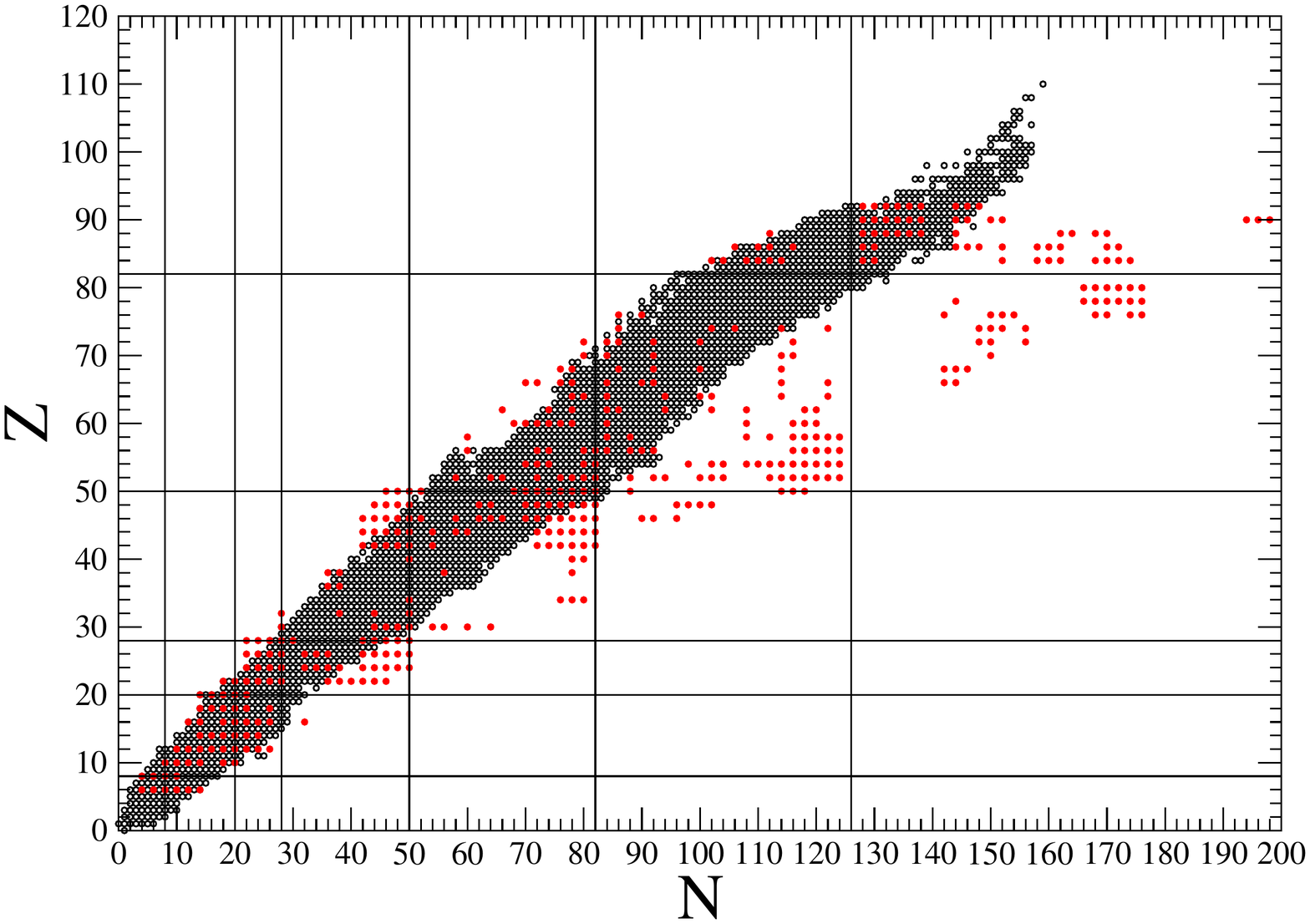}}
 \caption{(Color online) Microscopic axially-symmetric RHB prediction of nuclei that have small radial dispersion of the single-particle states of valence nucleons (red circles), plotted on the 
 background of empirically known nuclides in the $N- Z$ plane. The single-nucleon dispersions have been calculated using the functional DD-ME2 and separable pairing, 
and assuming axial symmetry.}    
 \label{fig:HOa}
\end{center}
\end{figure}

An interesting possibility however, is the formation of individual $\alpha$-like clusters from valence nucleons in heavy nuclei. We have performed a systematic, 
fully self-consistent Relativistic Hartree-Bogoliubov (RHB) \cite{vre05} calculation of single-nucleon dispersions in axially symmetric nuclei 
over the entire nuclear chart, using the functional DD-ME2. Pairing correlations 
have been taken into account by employing an interaction that is separable in 
momentum space, and is completely determined by two parameters adjusted 
to reproduce the empirical bell-shaped pairing gap in symmetric nuclear matter \cite{pairing}. The Dirac-Hartree-Bogoliubov equations are solved by expanding 
the nucleon spinors in a large axially-symmetric HO basis. 
The microscopic values of the dispersion $\Delta$r have been calculated for each single-particle state. Figure~\ref{fig:HOa} indicates, on the table of nuclides in the $N - Z$ plane, 
those nuclei in the RHB calculation for which both the neutron and proton valence states (having an occupation probability larger than 0.1) exhibit a significantly small dispersion, of the order of 1 fm. For deformed nuclei it can be shown 
that these Nilsson levels do originate from $n=1$ spherical states with the degeneracy raised by deformation. One notices that pronounced localization, as 
precondition for the formation of cluster structures, is present in light nuclei but also occurs among valence nucleons in medium-heavy and heavy nuclei, 
in agreement with empirically known alpha- and cluster-radioactive nuclei. For instance, favorable condition for clustering is predicted for $^{212}$Po, 
in accordance with experimental evidence \cite{ast10}. 
The EDF-based approach used in this work provides a global interpretation of the 
occurrence of cluster structures by means of spatial dispersion of single-nucleon wave functions.

\begin{figure}[h]
\scalebox{0.25}{\includegraphics{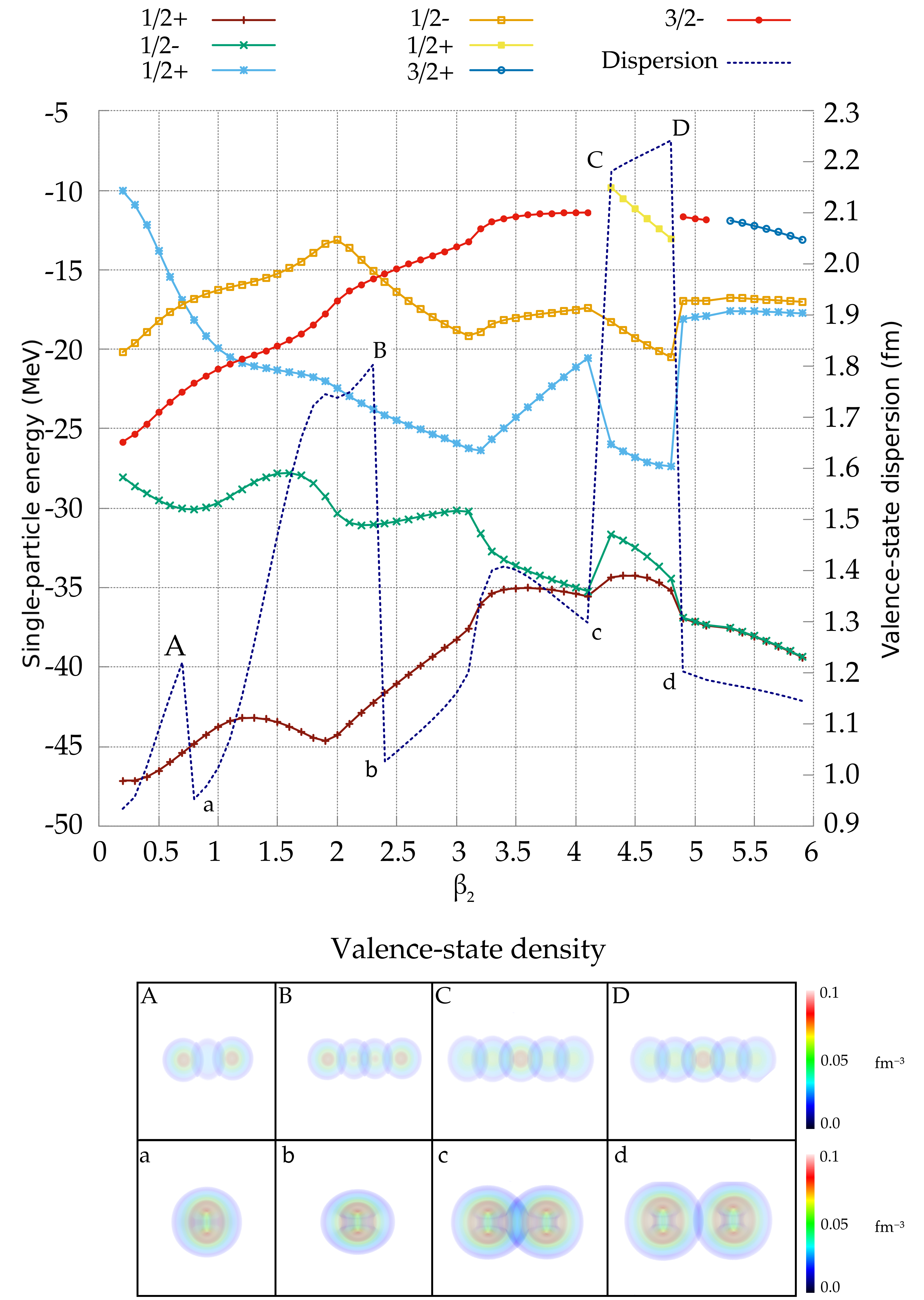}}
 \caption{(Color online) The occupied single-neutron levels $\Omega^\pi$ of $^{20}$Ne in the RMF approach as functions of the axial 
deformation parameter (solid curves), the dispersion of the wave function corresponding to the highest level occupied by the two valence neutrons 
(blue dotted curve), and the partial intrinsic 
densities of the valence neutrons for values of deformation that correspond to the peaks and minima of dispersion.}  
 \label{fig:me}
\end{figure}

The role of deformation, which is known to favor cluster formation \cite{ebr14,rae,zha16}, is illustrated in Fig.~\ref{fig:me} where we show the self-consistent 
mean-field results for $^{20}$Ne calculated using the relativistic density functional DD-ME2 in the RMF approach. 
Pairing does not play an important role for the effect that we consider in this particular nucleus, and it has not
been included in the RMF calculation restricted to axial symmetry.
Figure \ref{fig:me} displays the occupied single-neutron levels as functions of the axial 
deformation parameter, the dispersion of the wave function corresponding to the highest level occupied by the two valence neutrons, and the partial intrinsic 
densities of the valence neutrons for values of deformation that correspond to the peaks and minima of dispersion. In general, the spatial dispersion increases 
with deformation until a level crossing occurs for the last occupied state. The largest and sharpest increase of the spatial dispersion takes place at 
the deformation at which a $1/2^+$ state (originating from the $2s_{1/2}$ spherical state) becomes occupied. It is remarkable that at this point 
the dispersion increases by the factor $\sim$ 1.7, which we encountered above when discussing the filling of the $n=2$ HO state
(cf. Eq. (\ref{eq:d2}) and Fig.~\ref{fig:HOn}). The valence state partial densities exhibit more pronounced localization for small dispersion, 
while the largest spreading is obtained at the points C and D at which the 1/2$^+$ state becomes the last occupied neutron level. 

The present study can also be related to the discussion of (multi)clustering in super-deformed and hyper-deformed states \cite{rae,naza}. 
These specific states (ratio of deformed HO frequencies of 2 and 3, respectively) can be described as irreducible representations of SU(3). The magic numbers of super (hyper) deformed states are obtained from the sum of two (three) magic numbers of the spherical system. These relations involve small values of radial quantum numbers and, through Eq. (\ref{eq:res}), this can be linked to more localized states. The present approach however, as illustrated 
in Fig.~\ref{fig:me}, establishes a connection between spatial dispersion and clustering for all deformations, rather than only for specific super- and hyper-deformed 
states.

In summary, we have used the self-consistent mean-field framework based on nuclear energy density functionals, and the spherical harmonic oscillator approximation 
for the nuclear potential, to analyze the radial dispersion of single-nucleon wave functions. It has been shown that the dispersion exhibits a pronounced 
dependence on the radial quantum number, but essentially does not depend on the orbital angular momentum. In particular, for single-neutron states with $n=1$ the dispersion is of the size of the average inter-nucleon distance, and the correspondingly small value of the localization parameter $\alpha_{loc}$ indicates a transition between the nuclear quantum liquid phase and a hybrid phase of cluster states that coexists with mean-field states. Based on the 
HO approximation, we have derived an analytic expression for the localization parameter that, in addition to the dependence on the depth of the 
nuclear potential and the nucleon number, explicitly takes into account the radial quantum number of occupied single-nucleon states. 
For $A \leq 30$ and $n=1$ states occupied, $\alpha_{loc} \lesssim 1$ and the formation of $\alpha$-clusters is favored. Although in heavier nuclei levels that originate from $n > 1$ spherical states are largely delocalized and these systems exhibit the quantum liquid phase of nucleonic matter, individual 
 $\alpha$-like clusters can be formed from valence nucleons filling Nilsson levels that can be traced back to the $n=1$ spherical mean-field states. 
The role of deformation in the evolution of spatial dispersion of single-nucleon levels has been microscopically analyzed in the example of $^{20}$Ne, 
showing the robustness of the present analysis and conclusions. This study provides a general basis for understanding the conditions for cluster 
formation in light and heavy nuclei.

\bigskip
This work has been supported in part by the QuantiXLie Centre of Excellence, a project co-financed by the
Croatian Government and European Union through the European Regional
Development Fund - the Competitiveness and Cohesion Operational Programme
(KK.01.1.1.01).

\end{document}